\documentstyle[12pt]{article}
\title{Diffractive  light vector meson production at
large momentum transfers}

\author{D.Yu. Ivanov }
\date{Institute of Mathematics, 630090
Novosibirsk, Russia}

\begin{document}

\language=1

\maketitle

\begin{abstract}
The diffractive process $\gamma^*(Q^2) p\to V+X$ (where $V=
\rho^0, \omega , \phi$ are the vector mesons,
consisted of light quarks, $X$ represents the hadrons to that a
proton dissociates) is studied. We consider the region of large
momentum transfers, $|t|>>\Lambda^2_{QCD}$, and large energies,
s.  In the leading log approximation of perturbative QCD ( using
BFKL equation ) the asymptotic behaviour of the cross section in
the limit $s\to\infty , s>>|t|, Q^2$ is obtained.

We compare the results derived from BFKL equation with that
obtained in the lowest order of QCD (two--gluon exchange in the
$t$- channel).  The possibility to investigate these reactions
at HERA is discussed.
\end{abstract}

\section{Introduction}

This paper deals with the diffractive production of light vector
mesons ($V= \rho^0 ,\omega , \phi $) in $gamma^*(Q^2) p$
interactions
\begin{equation}
\gamma^* p\to V X \ ,
\label{1}
\end{equation}
where the meson $V$ and the hadrons $X$ to that the proton
dissociates are divided by the large rapidity gap.  $Q^2$ is the
virtuality of photon.  We will consider the process (\ref{1}) in
the region of large momentum transfers, $|t|>>\Lambda^2_{QCD}$,
and sufficiently large centre-of-mass energies, $s>>|t|, Q^2$.
The first condition justifies the using of perturbative QCD to
the description of the process (\ref{1}).  We reason that the
transferred momentum is large enough, therefore a hadron (VDM)
nature of photon will be neglected below.  The second condition
suggests the necessity of perturbative series summation, since
parameter $\alpha_s\ln{s}$ can be not small.

We will discuss both the diffractive photoproduction (process
(\ref{1}) at $Q^2=0$) and the diffractive production of mesons
in the deep inelastic scattering at arbitrary relation between
$Q^2$ and $|t|$.

Process (\ref{1}) has the vacuum quantum numbers in the $t-$
channel.  The summation of leading logarithms ($\alpha_s
\ln{(s)}$) in the vacuum channel leads to the famous BFKL
equation, that describes the perturbative ('hard') QCD pomeron
in the leading log approximation (LLA), see refs.
\cite{KLF,BL,L,Lpqcd}. Using the solution of BFKL equation, we
will obtain the asymptotic behaviour of the cross section of
the process considered at $s\to \infty$ and nonzero value of
momentum transfer.

At low momentum transfer the elastic process $\gamma^* p\to Vp$
was considered in refs. \cite{DonLan,Cud,Rys,KNNZ,BFGMS}.  The
interest to inelastic process (\ref{1}) relates with the
possibility to study BFKL pomeron in the modern experiments at
$ep$ collaider HERA in the wide range of parameters $s,|t|,Q^2$.

The process (\ref{1}) in the same kinematics was considered in
ref. \cite{ForRys}, where for the description of transition of
quark's pair to the meson, consisted of heavy quarks ( $V=
J/\Psi , \Upsilon$ ), the non--relativistic approximation was
used.  The process of light mesons production needs to be
considered separately, since here the non--relativistic
approximation is not applicable.  It is necessary to take into
account the nontrivial longitudinal motion of quarks in a light
meson in accordance with the standard leading twist approach to
the hard exclusive processes, see \cite{BrLe,ChZhit}.  Besides,
there is a substantial difference in the polarization state of
the meson produced in reaction (\ref{1}). For heavy mesons
production there is an approximate helisity conservation, the
helisity of meson coincides with the helisity of initial photon
$\lambda_M=\lambda_{\gamma}=\pm 1$, ÓÍ. \cite{GPS2}. In
contrast, the light mesons, as the consequence of a helisity
conservation at each vertex in the hiral limit of QCD, are
produced in a state with helisity $\lambda_M=0$, irrespective of
helisity state of initial photon.  The amplitudes of production
of light and heavy mesons have the different structure, compare
eqs.  (\ref{9},\ref{10}) and (\ref{11},\ref{12}).  The results
for amplitude of $J/\Psi$ production process cannot be directly
(by substitution $m_{J/\Psi}\to m_{\rho^0}$) transferred to
light meson production. Therefore, the conclusion of ref.
\cite{ForRys}
(which is based on this substitution), about a larger yield at
$|t|>>m^2_V$ of heavy vector mesons than that of the lighter
ones, seems to us as arbitrary one.

At sufficiently large momentum transfer ($-t>>\Lambda_{QCD}^2$)
the cross section of process (\ref{1}) may be represented as a
product of 'hard scattering' cross section and parton
distribution functions in a proton, see Fig.~1,
\begin{equation}
\frac{d\sigma (\gamma^* p\to VX)}{dtdx}=\left(
\frac{81}{16}G(x,t)+\sum_f\left[q(x,t)+\bar q(x,t)\right]\right)
\frac{d\sigma(\gamma^* q\to Vq)}{dt} \ .
\label{2}
\end{equation}
In ref. \cite{ForRys} some arguments are presented in a favor
that, at HERA energies and at $|t|\geq 2 \mbox{ çÜ÷}^2$ a
deviation from this simple picture ( the possibility to connect
$t-$ channel gluons with the different partons in proton ) does
not exceed $15\%$ for the events, in which the invariant mass of
the products of photon's disintegration is not too large,
$M^2_X\sim -t$.

In Born approximation (two--gluon exchange in the $t-$ channel)
the process of light vector meson production on quark
$\gamma^*(Q^2)q\to V q$ was considered in detail earlier
\cite{GIS}.
In the next section we obtain the results for this process in
LLA.  In the closing section the comparison of results of LLA
and Born approximation will be done, we also estimate the
possibility to study BFKL pomeron in this reaction at HERA. In
Appendix we will discuss the approximate method for the
calculation of amplitudes in LLA based on the result of
\cite{MuTa} for the amplitude of quark-quark scattering.

\section{Process $\gamma^* q\to Vq$ in LLA}

First we recall general LLA results of perturbative QCD and the
expressions which describe the photon to meson transition.  In
what follows we present results obtained for the process
considered separately for the case of real and virtual photon in
an initial state.

\subsection{General LLA expressions}

In LLA, as it was established in refs. \cite{KLF,BL,L}, the
amplitude of the (quasi)elastic scattering of two colorless
objects may be written as
\begin{eqnarray}
& A(s,t)=is\int\limits^{\sigma +i\infty}_{\sigma -i\infty} {d\omega
\over 2\pi i}({s \over \Lambda^2})^{\omega}f_\omega({\bf q}^2)\ ,
\quad t=-{\bf q}^2 \ , & \nonumber \\
& f_\omega(q^2)=\int
d^2{\bf k}_1d^2{\bf k}_2 \Phi^1({\bf k}_1,{\bf q})\Phi^2({\bf
k}_2,{\bf q}) f_\omega({\bf k}_1, {\bf k}_2, {\bf q})\ , &
\label{3}
\end{eqnarray}
where ${\bf k}_1,{\bf k}_2$ are the two-dimensional momenta of
the $t$- channel gluons in Fig. 1, \newline $f_\omega({\bf
k}_1,{\bf k}_2,{\bf q})$ is the partial wave amplitude of
gluon-gluon scattering in the color singlet channel, impact
factors $\Phi^{1,2}({\bf k},{\bf q})$ depend on the internal
structure of the scattering objects.  $\Lambda^2$ is the hard
scale of the process.  In the case when process (\ref{1})
is initiated by real photon, the only one hard scale is the momentum
transfer. Therefore, in this case $\Lambda^2\sim -t$.  Throughout
the paper all momentum vectors are two-dimensional vectors in the
plane orthogonal to the axis of reaction.

The impact--factors of colorless particles tend to zero when the
momenta of the gluons in the $t$- channel vanish
\begin{equation}
\Phi^{1,2}({\bf k},{\bf q})\bigm|_{{\bf k}=0}=
\Phi^{1,2}({\bf k},{\bf q})\bigm|_{{\bf k}-{\bf q}=0}=0 \ .
\label{4}
\end{equation}

Partial wave amplitude $f_\omega({\bf k}_1,{\bf k}_2,{\bf q})$
is the solution of BFKL integral equation \cite{KLF}.  In ref.
\cite{L} Lipatov has obtained the result for $f_\omega({\bf
k}_1,{\bf k}_2,{\bf q})$ in the important case, when the momentum
transfer is nonzero, ${\bf
q}\neq 0$. The high energy asymptotic of partial wave amplitude
can be written as
\begin{equation} f_\omega({\bf k}_1,
{\bf k}_2, {\bf q})= {1 \over 64\pi^6}\int {d\nu \over
(\nu^2+\frac{1}{4})^2} {\nu^2 \over
\omega
-\omega(0,\nu)} I^*_\nu({\bf k}_1,{\bf q})I_\nu({\bf k}_2,{\bf
q}) \ , \label{5}
\end{equation}
\begin{equation} I_\nu ({\bf
k},{\bf q})=\int d^2\rho_1d^2\rho_2 \left( {
({\mbox{\boldmath$\rho$}}_1 -{\mbox{\boldmath$\rho$}}_2)^2
\over \rho_1^2\rho_2^2 }
\right)^{{1\over 2}+i\nu} \cdot exp\left[i{\bf k}
{\mbox{\boldmath$\rho$}}_1 +
i({\bf q}-{\bf k}){\mbox{\boldmath$\rho$}}_2\right]  \
. \label{6} \end{equation}
At the derivation of esq. (\ref{5},\ref{6}) we keep only leading
($n=0$) term in eq. (29) of ref. \cite{L}, and perform the
simple rearrangements similar to those done in ref. \cite{MuTa}.

In ref. \cite{MuTa} the LLA result for the differential cross
section of quark--quark elastic scattering was derived.  In eq.
(\ref{5})
\begin{equation} \omega(n,\nu)={6\alpha_s \over
\pi}\chi(n,\nu)\ , \quad \chi(n,\nu)=Re\left[\psi(1)-
\psi\left({|n|+1 \over 2}+i\nu\right)\right] \ , \label{7}
\end{equation}
where $\psi $ is a standard logarithmic derivative of the Gamma
function.

It is necessary to define the impact--factors in
eq. (\ref{3}). Impact--factor of quark is
\begin{equation}
\Phi_{q\to q}=\alpha_s{\delta^{ab} \over N} \ ,
\label{8}
\end{equation}
here $N=3$ is the number of colors. Below, for the sake of
simplicity, we will discuss the production of $\rho^0$ meson,
the results for $\omega$ and $\psi$ production can be trivially
obtained by the change of coupling constants.  The
impact--factor of $\gamma\to \rho^0$ transition, that describes
the upper block of Fig.~1, was derived in ref. \cite{GPS1}
\begin{equation}
\Phi_{\gamma\to \rho^0}({\bf k},{\bf q})=
-e\alpha_s{\delta^{ab} \over 2N}f_\rho Q_\rho
\int\limits^{+1}_{-1}d\xi\varphi_\rho(\xi)\xi\left({\bf Q}{\bf e}
\right)  \ .
\label{9}
\end{equation}
According to the QCD leading twist approach to the hard
exclusive reactions
\cite{BrLe,ChZhit} this impact--factor has a form of a convolution
(the integral over $\xi$) of hard amplitude of photon to
collinear quark's pair transition and the process independent
meson's distribution amplitude $\varphi(\xi ,
\mu)$.
Variable $\xi$ describes the relative longitudinal motion of
quarks in a meson, $\mu$ is a factorization scale.  For the
photoproduction process it is natural to choose the value of
momentum transfer as the factorization scale, $\mu^2=-t$.  The
information about distribution amplitude can be obtained from
analysis of another exclusive processes or is derived by the
nonperturbative QCD methods (QCD sum rules, calculations on a
lattice).  Further we will use the simple parametrization for
$\rho^0$ meson distribution amplitude:
\begin{equation}
\varphi_\rho(\xi)={3 \over
4}(1-\xi^2)( 1-{b_V \over 5}+b_V\xi^2 ).
\label{fun}
\end{equation}
The value of parameter $b_V$ was estimated by the QCD sum rules
method in ref. \cite{ChZhit}.  At $\mu\simeq 0.5 \div 1.0 \mbox{
GeV}$, $b_V=1.5$.  The dependence of distribution amplitude on
the factorization scale is predictable in perturbative QCD, in
the lowest order
\begin{equation}
b_V(\mu_2)=b_V(\mu_1)\left[ \alpha_s(\mu_2)/\alpha_s(\mu_1)
\right]^{50/9b_0} \ , \quad b_0=11-{2\over 3}n_{fl} \ .
\label{evolv}
\end{equation}

In eq. (\ref{9}) $\delta^{ab}/N$ is the color factor, that
appears due to the projection on the colorless state of quark's
pair.  ${\bf e}$ is the polarization vector of photon,
$f_\rho=210\mbox{ MeV}$ is the dimensional coupling constant.
$Q_\rho=1/\sqrt{2}$ is connected with the charge content of
$\rho^0$ meson, $|\rho^0 >=(|u\bar u > - |d\bar d>)/\sqrt{2}$.
Vector ${\bf Q}$ in eq. (\ref{9}) has the form
\begin{equation}
{\bf Q}=
\left[
{ {\bf q} \over {\bf q}^2{1+\xi \over 2} } +
{ {\bf k}-{\bf q}{1+\xi \over 2}
\over ({\bf k}-{\bf q}{1+\xi \over 2})^2 }
\right] -
[\xi \leftrightarrow - \xi ] \ .
\label{10}
\end{equation}
The terms that do not depend on ${\bf k}$ in eq. (\ref{10})
describe that diagrams as depicted in Fig.~1, when $t-$ channel
gluons link to the same quark (or antiquark) line at the upper
block of Fig.~1.  Depending on ${\bf k}$ terms in eq.~(\ref{10})
describe the diagrams where the qluons link to various quark's
lines.

Note, that eq.~(\ref{9}) describes the transition of photon to
longitudinal polarized meson. The impact--factor of production
of meson in a state with helisity $\lambda=\pm 1$ is
\begin{equation}
\Phi_{\gamma\to \rho^0}({\bf k},{\bf q})=
e\alpha_s{\delta^{ab} \over 2N}f_\rho Q_\rho
\int\limits^{+1}_{-1}d\xi\varphi_\rho(\xi)mR\left({\bf e}{\bf
e^*_M}\right) \ ,
\label{11}
\end{equation}
here $m$ is a quark mass of which the meson consists, $\bf e_M$
is a polarization vector of meson. Scalar $R$ is symmetric about
the permutation of quark's momenta
\begin{equation}
R=
\left[
{ 1 \over \left({\bf q}{1+\xi \over 2}\right)^2 } +
{ 1 \over ({\bf k}-{\bf q}{1+\xi \over 2})^2 }
\right] +
[\xi \leftrightarrow - \xi ] \ .
\label{12}
\end{equation}
To produce the meson in a state with helisity $\lambda=\pm 1$ it
is necessary to flip the quark's spin. It is the reason why
amplitude (\ref{11}) is proportional to $m$, the current mass of
quark.  Therefore, the light mesons can be produced in a state
with helisity $\lambda=0$ only.

In high energy limit $\alpha_s\ln{(s)}>>1$, the region of small
$\nu$ gives the main contribution to the amplitude in
eqs.~(\ref{5},\ref{6}). Therefore we can integrate over $\nu$
and represent amplitude in the factorize form
\begin{equation}
A(s,t)=is{\exp{(\rho\ln{4})} \over \pi
\left( 28\pi^3\zeta (3)\rho \right)^{3/2}}
J_2^*(q)J_1(q) \ ; \quad \rho={6\alpha_s \over \pi}
\ln{(s/\Lambda^2)} \ .
\label{13}
\end{equation}

All dependence on the structure of particles participating in
reaction contains in the factors $J_{1,2}$
\begin{equation}
J_{1,2}(q)=\int d^2k \Phi^{1,2}({\bf k},{\bf q})I_{\nu=0}({\bf
k},{\bf q}) \ .
\label{14}
\end{equation}

Factor $J_2(q)$ appears at the calculation of quark--quark
scattering in LLA, it was found in ref.~ \cite{MuTa}
\begin{equation}
J_2(q)=-{2\alpha_s \over 3}{(2\pi )^3 \over q}\delta^{ab} \ .
\label{15}
\end{equation}

\subsection{Real photon in initial state.}

Let us define factor $J_1(q)$ that describes upper part of
Fig.~1, the transition of $\gamma \to \rho^0$. The integration
over $d^2k$ in eq.~(\ref{14}) gives
\begin{eqnarray}
& \int d^2k \exp{(i{\bf k}{\mbox{\boldmath$\rho$}})}
\Phi_{\gamma\to\rho^0} ({\bf k},{\mbox{\boldmath$\rho$}})=
-\frac{\displaystyle e\alpha_s}{\displaystyle 6}
\delta^{ab}Q_{\rho}f_\rho\int\limits^{+1}_{-1}d\xi
\xi\varphi_\rho(\xi )
\times & \nonumber \\
&
\left[ (2\pi )^2
        \left(\frac{\displaystyle -4\xi }{\displaystyle 1-\xi^2}
    \right)
\delta^2({\mbox{\boldmath$\rho$}})
\frac{\displaystyle {\bf e}{\bf q}}{\displaystyle  q^2 }+(2\pi
i) \frac{\displaystyle {\bf
e}{\mbox{\boldmath$\rho$}}}{\displaystyle \rho^2}
 \left\{ \exp{(i{\bf q}{\mbox{\boldmath$\rho$}}{1+\xi \over 2})}
 - \exp{(i{\bf q}{\mbox{\boldmath$\rho$}}
 {1-\xi \over 2})}\right\} \right] \ . &
 \label{16}
\end{eqnarray}
Note, that the contribution of the first $\sim
\delta^2({\mbox{\boldmath$\rho$}})$ term in righthand side of
eq.~(\ref{16}) vanishes at the subsequent integration over
variables $\rho_1$ and $\rho_2$. The contribution of the second
term
\begin{equation}
J_1(q)=-(4\pi i){e\alpha_s \over 6}
\delta^{ab}Q_{\rho}f_\rho\int\limits^{+1}_{-1}\xi\varphi_\rho(\xi)
d\xi \int d^2\rho_1d^2\rho_2
{ {\bf e}({\mbox{\boldmath$\rho$}}_1-{\mbox{\boldmath$\rho$}}_2)
\over |{\mbox{\boldmath$\rho$}}_1||{\mbox{\boldmath$\rho$}}_2|
|{\mbox{\boldmath$\rho$}}_1-{\mbox{\boldmath$\rho$}}_2| }
e^{i{\bf q}({\mbox{\boldmath$\rho$}}_1{1+\xi \over 2}+
{\mbox{\boldmath$\rho$}}_2{1-\xi \over
2})} \ .
\label{17}
\end{equation}
Let us next write $J_1(q)$ as
\begin{equation}
J_1(q)=-(16\pi ){e\alpha_s \over 6q^2}
\delta^{ab}Q_{\rho}f_\rho({\bf e}{\bf n})
\int\limits^{+1}_{-1}\xi\varphi_\rho(\xi)d\xi
{d \over d\xi}I(\xi) \ ; \quad {\bf n}={ {\bf q}\over q} \ .
\label{18}
\end{equation}
Where function $I(\xi)$ is the integral over dimensionless
vectors ${\bf r}_{1,2}$
\begin{equation}
I(\xi)=\int { d^2{\bf r}_1d^2{\bf r}_2 \over
|{\bf r}_1+{\bf r}_2||{\bf r}_1-{\bf r}_2|r_2 }
e^{ i{\bf n}({\bf r}_1+\xi{\bf r}_2) } \ .
\label{19}
\end{equation}
After the simple  rearrangements we obtain
\begin{equation}
I(\xi)=2\pi \int
{ d^2{\bf r} \over |{\bf r}+{\bf n}||{\bf r}-{\bf n}||{\bf r}-\xi{\bf
n}| } \ .
\label{20}
\end{equation}
Unfortunately, we have failed to obtain the analytical expression for
$I(\xi )$. Therefore, in eq.~(\ref{18})
the integral over $\xi$
with the distribution amplitude  of a meson
was   calculated  firstly
 and then the integration over
$d^2{\bf r}$ was performed.
\footnote{At this stage a transition to the
elliptic coordinate was used.   }
As a result we obtain
\begin{equation}
J_1(q)=-{\pi^5 \over 2} {e\alpha_s \delta^{ab}Q_\rho f_\rho({\bf
e}{\bf n}) \over  q^2 }\eta_V \ ; \quad
\eta_V=(1+b_V{131 \over 320}) \ .
\label{21}
\end{equation}

Eqs.~(\ref{13},\ref{15},\ref{21}) define the amplitude of the meson
photoproduction on quark, $\gamma q \to \rho^0 q$.
The cross section of this reaction may be expressed in terms of
experimental
measured quantity --- the width of decay of
$\rho^0$ meson into  $e^+e^-$ pair $\Gamma^V_{e^+e^-}={4\pi
\alpha^2f_V^2Q^2_V \over 3m_V}$,
\begin{equation}
{d\sigma^{BFKL} \over dt}={4\pi^4\eta_V^2\over 1029\zeta^3(3)}
{\alpha_s^4m_V\Gamma^V_{e^+e^-} \over \alpha |t|^3}
{e^{(2\rho\ln{4})} \over \rho^3 }({\bf en})^2 \ .
\label{22}
\end{equation}

\subsection{Virtual photon in initial state.}

In previous subsection we have obtained
the analytical results for the process (\ref{1})
in a case of photoproduction, when $Q^2=0$. Let us go to the
consideration of mesons production by off shell photons.

Note again, that discussed above helisity conservation holds for
the both real and virtual photons in initial state.  The
perturbative QCD thus predicts the definite helisity of the
produced mesons, $\lambda=0$.

The impact--factors of $\gamma^*(Q^2)\to \rho^0$ transition were
derived earlier \cite{GIS}.  The off shell photon can be
longitudinal ($S$) or transversely ($T$) polarized.  First we
consider a process initiated by $T$- photon.  The impact factor
of $\gamma^{(T)}\to \rho^0$ transition has the same form as the
impact--factor for real photon (\ref{9}).  The inclusion of
photon's virtuality leads to modification of structure ${\bf
Q}$,
\begin{equation}
{\bf Q}\to {\bf Q}=
\left[
{ {\bf q}{1+\xi \over 2} \over
{1-\xi^2\over 4}Q^2 + \left({\bf q}{1+\xi \over 2}\right)^2 } +
{ {\bf k}-{\bf q}{1+\xi \over 2}
\over {1-\xi^2\over 4}Q^2 + ({\bf k}-{\bf q}{1+\xi \over 2})^2 }
\right] -
[\xi \leftrightarrow - \xi ] \ .
\label{23}
\end{equation}
Now factor $J_1$ depends on two variable (on momentum transfer
$q$ and on virtuality of photon $Q$). After the simple
rearrangements it can be brought to the form
\begin{equation}
J_1^{(T)}=-(16\pi ){e\alpha_s \over 6q^2}
\delta^{ab}Q_{\rho}f_\rho({\bf e}{\bf n})
\int\limits^{+1}_{-1}\xi\varphi_\rho(\xi)
L^{(T)}(\eta , \xi) d\xi \ ;
\quad {\eta}={ Q\over q}\sqrt{(1-\xi^2)} \ ;
\label{24}
\end{equation}
here $L^{(T)}(\eta , \xi)$ is expressed in terms of Bessel
functions of the first kind $J_1$ É $K_1$:
\begin{equation}
L^{(T)}(\eta , \xi)=\int { d^2k_1 d^2k_2 ({\bf k}_1{\bf k}_2) \over
k_1 |{\bf k}_1+{\bf k}_2(1+\xi )| |{\bf k}_1-{\bf k}_2(1-\xi )| }
J_1(k_1)\eta K_1(\eta k_2) \ .
\label{25}
\end{equation}

It can be shown that in the limit $Q^2 << q^2$ the obtained
result  reduces to
that one derived in the
previous subsection for real photon, $L^{(T)}(\eta ,\xi)
\to {d \over d\xi}I(\xi)$, see eqs.~(\ref{18},
\ref{19}, \ref{21}).

When the virtuality of photon is high $Q^2 >> q^2$, the main
contribution to integral (\ref{25}) originates from the region
of small $k_2$, $k_2\leq 1/\eta$. Therefore
\begin{eqnarray}
& L^{(T)}(\eta , \xi)\left|_{\eta >>1}\right. \simeq
\int \frac{\displaystyle d^2k_1 d^2k_2 ({\bf k}_1{\bf k}_2)}
{\displaystyle k_1^3
\left(1+2\xi { ({\bf k}_1{\bf k}_2) \over k_1^2 } \right) }
J_1(k_1)\eta K_1(\eta k_2) \simeq & \nonumber \\
& -2\pi\xi\int \frac{\displaystyle k_2^2 dk_1 d^2k_2}
{\displaystyle k_1^2 } J_1(k_1)\eta K_1(\eta k_2) \ . &
\label{26}
\end{eqnarray}
This expression is logarithmically divergent at the lower limit
of the integration on $k_1$. This singularity slows down at
$k_1\sim 1/\eta$.  Hence, the result with the logarithmic
accuracy can be written as
\begin{equation}
L^{(T)}(\eta , \xi)\left|_{\eta >>1}\right. \simeq
-\pi\xi\ln{(\eta )}\int d^2k_2k^2_2
\eta K_1(\eta k_2) = -3\pi^3\xi {\ln{(\eta )}  \over \eta^3 } \ .
\label{27}
\end{equation}
Performing the integration over $\xi$ in eq.~(\ref{24}) we obtain
\begin{eqnarray}
&
\left. J_1^{(T)} \right|_{Q^2>>q^2}\approx 3\pi^5e\alpha_s
\delta^{ab}\frac{\displaystyle Q_{\rho}f_\rho({\bf e}{\bf n})q}{
\displaystyle Q^3}\times & \nonumber \\
& \left[ \ln{{Q\over q}}(1+{11\over 20}b_V) -{(1+\ln{4})\over 2}-
{b_V\over 2}({74\over 80}+{44\over 80}\ln{4}) \right] \ . &
\label{27new}
\end{eqnarray}

Let us turn to the case of longitudinal polarization of initial
photon.  The impact--factor of $\gamma^{(S)}\to \rho^0$
transition has the form
\begin{equation}
\Phi_{\gamma\to \rho^0}({\bf k},{\bf q})=
-e\alpha_s{\delta^{ab} \over 2N}f_\rho Q_\rho
\int\limits^{+1}_{-1}d\xi\varphi_\rho(\xi)
{1-\xi^2\over 2}\sqrt{Q^2}R(Q) \ .
\label{28}
\end{equation}
The amplitude of (S) photon to meson transition is proportional
to the virtuality of photon. Besides, the dependence on the
photon's virtuality is contained in structure $R(Q)$
\begin{equation}
 R(Q)=
\left[
{ 1 \over
m^2+{1-\xi^2\over 4}Q^2 + \left({\bf q}{1+\xi \over 2}\right)^2 } -
{ 1
\over m^2+{1-\xi^2\over 4}Q^2 + ({\bf k}-{\bf q}{1+\xi \over 2})^2 }
\right] +
[\xi \leftrightarrow - \xi ] \ .
\label{29}
\end{equation}
For the subsequent discussion of the heavy mesons production we
keep the quark mass in eq.~(\ref{29}).  Factor $J_1$ for the
scalar photon is
\begin{equation}
J_1^{(S)}=(16\pi ){e\alpha_sQ \over 6q^3}
\delta^{ab}Q_{\rho}f_\rho
\int\limits^{+1}_{-1}(1-\xi^2)\varphi_\rho(\xi)
L^{(S)}(\eta , \xi) d\xi \ ,
\label{30}
\end{equation}
where $L^{(S)}(\eta , \xi)$ is expressed in terms of Bessel
functions of the zeroth kind
\begin{equation}
L^{(S)}(\eta , \xi)=\int d^2k_1 d^2k_2
{ k_2 J_0(k_1) K_0(\eta k_2) \over
 |{\bf k}_1+{\bf k}_2(1+\xi )| |{\bf k}_1-{\bf k}_2(1-\xi )| }
 \ .
\label{31}
\end{equation}
In the limiting case $Q^2>>q^2$
\begin{equation}
L^{(S)}(\eta , \xi)\left|_{\eta >>1}\right. \simeq
 2\pi^3 {\ln{(\eta )}  \over \eta^3 } \ .
\label{32}
\end{equation}
Substituting (\ref{32}) in eq.~(\ref{30}) and performing the
integration over $\xi$ we obtain
\begin{equation}
\left. J_1^{(S)}(q)\right|_{Q^2>>q^2}\approx 2\pi^5e\alpha_s
\delta^{ab}{Q_{\rho}f_\rho \over Q^2}
\left[ \ln{{Q\over q}}(1+{b_V\over 20}) +{(1-\ln{4})\over 2}-
b_V({3\over 80}+{\ln{4}\over 40}) \right] \ .
\label{32new}
\end{equation}

The asymptotic expression (\ref{32new}) for the amplitude of a
process initiated by (S) photon becomes valid at lower
virtuality of photon (at fixed $q$) in comparison with $Q^2$ for
that the asymptotic expression (\ref{27new}) for the amplitude
of a process initiated by (T) photon comes into play. Compared
to impact--factor of $\gamma^{(T)}\to \rho^0$ transition, the
impact--factor of $\gamma^{(S)}\to
\rho^0$  contains  additional factor $1-\xi^2$,
which at the subsequent integration over $\xi$ in eq.~(\ref{30})
suppresses the contributions of the end point regions $\xi\to
\pm 1$.
Thus the mean $\eta$ at the integration in eq.~(\ref{30}) is
$\sim Q/q$ and the constant, which in eq.~(\ref{32new}) is
subtracted from the term $\sim \ln{({Q\over q})}$, is small.
The same constant in the asymptotic expression (\ref{27new}) for
the amplitude of process initiated by (T) photon is
substantially larger.

In ref. \cite{ForRys} integral (\ref{31}) was calculated
numerically at $\xi=0$ for various values of $\eta$. It turns
out that asymptotic expression (\ref{32}) differs from the
numerical result less than on 10\% when $\eta \geq 3 $.
Therefore, we will anticipate that asymptotic expression
(\ref{32new}) can give the result that is closed to true one
when $Q^2/q^2\geq 10$, in so far as the end point regions (when
$\xi\to \pm 1$) do not give the substantial contribution at the
subsequent integration over $\xi$ in eq.~(\ref{30}). Further we
will use equation (\ref{32new}) for the estimates of cross
section at the HERA energies. The inaccuracy connected with
using herewith asymptotic eq.~(\ref{32new}) rather than exact
results (\ref{30},\ref{31}) can not exceed 10\% when
$Q^2/q^2\geq 10$.

\section{Discussion}

In the previous part we derive in LLA the cross section of
vector meson photoproduction on quark at $s/|t|\to \infty$. Let
us compare (\ref{22}) with the obtained in ref.~\cite{GPS1}
result for the cross section in Born approximation (two--gluon
exchange in the $t-$ channel)
\begin{equation}
{d\sigma^{2G} \over dt}={64\pi\over 3}
{\alpha_s^4m_V\Gamma^V_{e^+e^-} \over \alpha |t|^3}
({\bf en})^2\upsilon_V^2\, ;\quad \upsilon_V=
(1+7/15b_V) \ .
\label{33}
\end{equation}

In  Born approximation the cross section does not depend on
the collision energy. In LLA the cross section increases as a
power of energy. Nevertheless, its ratio
\begin{equation}
{d\sigma^{BFKL} \over d\sigma^{2G}}=
{\pi^3\over 5488\zeta^3(3)}{\eta^2_V \over \upsilon^2_V}
{\exp{(2\rho\ln{4})} \over \rho^3}\simeq
0.003\cdot {\exp{(2\rho\ln{4})} \over \rho^3}
\label{34}
\end{equation}
exceeds a unit only at the sufficiently large values of parameter
$\rho={6\alpha_s  \over \pi }\ln{(s/\Lambda^2)} $,
when $\rho\geq 3.43$.
For the process (\ref{1}) at $Q^2=0$ it is natural to choose
the square of momentum transfer as
$\Lambda^2$ and the normalization point of strong interactions
constant  $\Lambda^2={\bf q}^2$,
$\alpha_s({\bf q}^2)$.  In  Table we give the values of (CM)
energies at that the ratio (\ref{34}) is equal to unit
depending on momentum transfer:

\begin{center}
\begin{tabular}{|c|c|c|c|c|c|c|c|c|} \hline
${\bf q}  (\mbox{ GeV }) $ & 3 & 4 & 5 & 6 & 7 \\ \hline
$\sqrt{s}(\mbox{ GeV })  $& 104 & 201 & 335 & 510 & 720  \\ \hline
\end{tabular}
\end{center}

In the region of large momentum transfers, $q\geq 4\div 5\mbox{
GeV}$, high energy asymptotic of LLA exceeds Born predictions
only when $\sqrt{s_{\gamma p}} > 200 \div 340 \mbox{ GeV}$.
These energies are too high even for HERA collaider. At these
transfers and energies available at HERA $\sqrt{s_{\gamma p}}
\sim 50\div 200
\mbox{ GeV}$, the expressions obtained in the high energy limit of
LLA are not applicable. Since they give the result which is less
than that obtained in a lowest order. In that instance the more
accurate analysis of BFKL series is needed.  One can try to
calculate a few first items of the expansion in $\alpha_s$, by
means of iterating of BFKL equation for partial amplitude.  In
the region under consideration the parameter $\rho$ is
sufficiently large, $\rho\sim 2\div 3.4$.  Since the required
number of iterations is $\sim\rho$, this analysis will be
complicated.

Nevertheless, it is natural in our opinion that Born result can
give a lower bound for the cross section.  The value of cross
section obtained in the two--gluon approximation is not too
small. According to eqs.~(\ref{2},\ref{33}) when $x>0.01$ and
$|t|\geq 16\mbox{ GeV}^2$ the total cross section of hard
diffractive production of $\rho^0 $ meson $\sigma^{2G}(\gamma
p\to \rho^0 X)\approx 4.6\mbox{ nb}$, when $x>0.01$ and $|t|\geq
25\mbox{ GeV}^2$ this cross section reduces to $\approx
1.1\mbox{ nb}$.
\footnote{At the integration of the differential cross section
(\ref{33}) the dependence of $\alpha_s$ on transfer momentum was
taken into account.}

At not too large transfers, $q\sim 3\div 4\mbox{ GeV}$, the
energy region, where BFKL result is larger than Born one,
$\sqrt{s_{\gamma p}} > 100 \div 200 \mbox{ GeV}$, is available
at HERA.  However these transfers most likely are not
sufficiently high for applicability of perturbative QCD to the
description of photon to meson transition. In the paper
\cite{GIS} some arguments are done in
a favor that the boundary value of momentum transfer, from which
the asymptotic (in $1/q$) equations of perturbative QCD
(\ref{9},\ref{11}) become applicable, is $\sim 4\div 5\mbox{
GeV}$ for Born amplitude. \footnote{ The Born amplitude gets the
sizable contribution on the regions where one of the virtual
quarks at the upper block of Fig.~1 is near mass shell. For the
self--consistency of the perturbative approach it is necessary
that the relative value of the contribution of nonperturbative
regions, where the virtuality of quark $\leq
\Lambda^2_{QCD}$, was small.
According to ref.~\cite{GIS}, this criteria is fulfilled
beginning with sufficiently large transfer momenta $q\geq 4\div
5\mbox{ GeV}$.}

Investigating processes (\ref{1}) over a wide range of
transfers, it would be possible to determine the range of
validity of perturbative QCD.  Important information about the
region of validity of perturbative QCD can give the measurement
of meson's polarization.  Recall that perturbative QCD predicts
the definite helisity of produced meson, $\lambda=0$. To the
contrary, in the region of small transfers one would expect, in
the spirit of the VDM model, a helisity conservation in the
transition $\gamma^*\to \rho^0$.

We show that in the process of hard diffractive photoproduction
of light vector mesons the BFKL behaviour is attained only at
the sufficiently large values of $s/t$, see Table. The origin of
anomalously small numerical coefficient ($\sim 0.003$) in
equation (\ref{34}) is connected {\it mainly} with the small
numerical coefficient in general equation for amplitude in high
energy limit of LLA, see eq.~(\ref{13}). In addition, it is
known
\cite{MuTa,ForRys,RLaK0}, that high energy asymptotic of the solution of
BFKL equation displays a smaller degree of infrared singularity
as compared with one of sum of diagrams that describe the
process in the lowest order.  This leads to appearance of
additional infrared logarithm $\ln{q^2_1/q^2_2}$ at the
integration of Born amplitude over $d^2{\bf k}$ (over the
momenta of gluons in the $t-$ channel), where $q_{1,2}$ are the
quark's momenta at Fig.~1. This additional logarithm leads at
the integration over relative longitudinal momentum of quarks to
larger numerical coefficient in Born amplitude as compared to
that in the LLA amplitude.

Let us discuss the diffractive production of $\rho^0$ mesons
initiated by off shell photon. In the previous section we
derived expressions for the amplitudes in terms of integrals:
(\ref{24},\ref{25}) for transversal polarization, and
(\ref{30},\ref{31}) for longitudinal polarization of photon. In
fact, the amplitudes are determined by three--dimensional
integrals and can be evaluated numerically.  But this
calculation is difficult enough, because the integrand
expression oscillates. Instead, we will analyze the process
(\ref{1}) using the obtained above analytical expressions
(\ref{27new},\ref{32new}) for the asymptotic of the amplitudes
in the limit $Q^2>>q^2$.

Let us compare LLA asymptotic expressions and ones obtained in
Born approximation: \newline {\it for transversal polarization
of photon}
\begin{equation}
\left. {d\sigma^{BFKL} \over d\sigma^{2G}} \right|_{Q^2>>q^2}\approx
0.03\cdot {Q^2\over q^2\ln{(Q/q)}^2}{\exp{(2\rho\ln{4})} \over \rho^3}
\label{35} \ ;
\end{equation}
\newline {\it for longitudinal polarization of photon}
\begin{equation}
\left. {d\sigma^{BFKL} \over d\sigma^{2G}} \right|_{Q^2>>q^2}
\approx
0.01\cdot{Q^2\over q^2}{\exp{(2\rho\ln{4})} \over \rho^3}  \ .
\label{36}
\end{equation}
Composing relations (\ref{35},\ref{36}), we use results of
ref.~\cite{GIS} for the amplitude of process
$\gamma^*(Q^2)q\to \rho^0q$ in Born approximation.

In the limit of large virtualities of photon the asymptotic
result of LLA for the amplitude exceeds substantially that
obtained in the two--gluon approximation.  Note the additional
enhancing factor $Q^2/q^2$ in eqs.~(\ref{35},\ref{36}) in
comparison with eq.~(\ref{34}).  When $Q^2>>q^2$, relevant scale
of the process is determined in Born approximation by the
transverse distance between quarks in $q\bar q$ pair with which
the two--gluon system is connected. This distance $\rho\sim 1/Q$
and does not depend essentially on the value of momentum
transfer $q$. BFKL amplitude, at $Q^2>>q^2$, is characterized by
two different scales: by the size of quark's pair $\rho_1\sim
1/Q$, and by the intrinsic distance between gluons in BFKL
ladder $\rho_2\sim 1/q$. In the course of development of gluon
ladder the transition from scale $\sim 1/Q$ to larger one $\sim
1/q$ occurs. Therefore, BFKL amplitude, as distinct from Born
one, depends on both virtuality of photon and momentum transfer
at $Q^2>>q^2$.

Let us estimate the value of cross section for process (\ref{1})
integrated over the region of transfers $2\mbox{ GeV}^2\leq q^2
\leq 3\mbox{ GeV}^2$ at $Q^2=25\mbox{ GeV}^2$ and HERA energies.
First of all note, that in this region the amplitude of process
initiated by scalar photon $A^{(S)}$ is larger than that one
initiated by transversely polarized photon
$A^{(T)}$.\footnote{In the Born approximation
\cite{GIS} these amplitudes coincide at $Q^2=q^2$. At the further
growth of photon's virtuality the amplitude $A^{(S)}$ decreases
more slowly than $A^{(T)}$.} Therefore we will estimate the
value of cross section of process (\ref{1}) for the case of
longitudinal polarization of initial photon.  We believe that in
the considered region (where $Q^2/q^2\sim 10$) asymptotic
expression (\ref{32new}) for the amplitude will give the result,
differing from exact one by the value which does not exceed
$10\div 20\%$.  According to
eqs.~(\ref{32new},\ref{13},\ref{15})
\begin{equation}
\left.{ d\sigma^{BFKL}_{\gamma^{(S)}q\to \rho^0 q} \over
dt }\right|_{Q^2/q^2>>1}\approx
0.96\cdot{\alpha^4_s\over \alpha}
{m_V \Gamma^V_{e^+e^-}\over Q^4q^2}
\left[ \ln{{Q^2\over q^2}} -0.505\right]^2
{\exp{(2\rho\ln{4})} \over \rho^3} \ .
\label{40}
\end{equation}
The dependence of the amplitude of process $\gamma^{(S)}q\to
\rho^0 q$ on
the shape of meson's distribution amplitude is not too strong.
Therefore we present result (\ref{40}) for the distribution
amplitude with parameter $b_V=1.0$. According to
eq.~(\ref{evolv}), this value of $b_V$ is accepted when $\mu
\sim 3\div 5
\mbox{ GeV}$.

Contrary to discussed above photoproduction process, where it is
natural to choose as the normalization point of strong
interaction constant the transferred momentum $q$ as well as to
put $\Lambda^2=q^2$, there are two scales ($Q^2=25 \mbox{
GeV}^2$ and $q^2\sim 2\div 3 \mbox{ GeV}^2$) in the kinematical
region under consideration.  If to choose as the hard scale of
the process $\Lambda^2=Q^2$ and to use the same value as the
argument of $\alpha_s$ in eq.~(\ref{40}) then integrated over
the region $2\mbox{ GeV}^2\leq q^2
\leq 3\mbox{ GeV}^2$, $x\geq 0.1$
cross section $\sigma^{BFKL}(\gamma^{(S)} (Q^2=25\mbox{ GeV}^2)
p\to \rho^0 X)$ constitutes $\approx 0.19\mbox{ nb}$ at
$\sqrt{s_{\gamma p}}=100 \mbox{ GeV}$, and $\approx 0.48\mbox{
nb}$ at $\sqrt{s_{\gamma p}}=200 \mbox{ GeV}$.

Point out that the results of calculation are extremely
sensitive to the choice of normalization point for $\alpha_s$
and $\Lambda^2$ in eq.~(\ref{40}). If as $\Lambda^2$ and
normalization point for strong interaction constant to take
$0.8\cdot Q^2$, then the results for cross sections increase
more that 1.5 times. This uncertainty is connected with the
using of LLA. Unfortunately, the results for QCD pomeron in the
next to leading logarithm approximation are not derived up to
now.  However the work in this way is in progress now, see
\cite{Fadin1,Fadin2,Fadin3}.
Therefore, in the complicated case, when there are two strongly
differing scales ($1/Q$ and $1/q$), we can not do a justified
choice of normalization point for strong constant and parameter
$\Lambda^2$.  Estimating cross section in the previous
paragraph, we choose the lowest scale, $1/Q$. We hope that
obtained therewith numbers are the estimates for cross section
from below.

Obtained above estimates for cross sections are not too small.
Therefore there is a hope to investigate the process (\ref{1})
at HERA in the region of large photon's virtualities
$Q^2>>q^2\sim 2\div 3\mbox{ GeV}^2$. In this region we predict
the fast increase of cross section with the energy growth of
$\gamma^* p$ system.  At $Q^2=25 \mbox{ GeV}^2$ and $q^2\sim
2\div 3 \mbox{ GeV}^2$ we expect the growth of cross section
more than 2.5 times at the increasing of $\sqrt{s_{\gamma p}}$
from $100$ to $200\mbox{ GeV}$.  It will be interesting to
handle experimental data using eqs.~(\ref{40},\ref{2}), varying
herewith the normalization point of $\alpha_s$ and parameter
$\Lambda^2$.

\vspace{1cm}

{\it \bf In conclusion.}

\vspace{0.7cm}

In the high energy limit of LLA the process of diffractive light
vector meson production was studied in the region of large
momentum transfers at the arbitrary virtuality of initial
photon. In an analytical form the results for photoproduction
(when $Q^2=0$) and for the limiting case of large virtuality
($Q^2>>q^2$) are derived.

It turns out that the LLA amplitude of process (\ref{1}) does
not exceed at $Q^2=0$ the Born one even at the energies of
collider HERA.

When $Q^2>>q^2>>\Lambda^2_{QCD}$ the LLA result is larger then
the Born one. The corresponding cross section therewith is not
too small. In this region we expect the fast, typical for BFKL
pomeron, growth of the cross section of process (\ref{1}) with
the energy of $\gamma^* p$ collisions.

\vspace{1cm}

{\bf Acknowledgments.} I am thankful to I.F.~Ginzburg and
A.P.~Burichenko for useful discussions. This work was supported
by a International Science Foundation (Grant RPL300) and a
Russian Foundation of Fundamental Research (Grant
93--02--03832).

\section*{Appendix}

Inserting (\ref{11},\ref{29}) in eq.~(\ref{13}) we obtain the
amplitude of hard diffractive production of the mesons consisted
of heavy quarks, $J/\Psi$, $\Upsilon$. The distribution
amplitude of heavy meson has a small width $\sim \upsilon/c<<1$,
where $\upsilon$ is a velocity of quarks in the meson. If the
approximation $\varphi (\xi)=\delta(\xi)$ is used, the
calculation is simplified drastically and reduced to the
discussed above case of meson production by scalar photon, see
eqs.~(\ref{9},\ref{11}). Relevant results may be obtained by the
simple replacement in eqs.~(\ref{30},\ref{31}): $-{\sqrt{Q^2}
\over 2} \to m({\bf ee}^*_M)$
and $ {\eta}\to \eta^\prime =\sqrt{4m^2 + Q^2 \over q^2}$. They
are in agreement with that obtained in ref. \cite{ForRys}.

At the same time, the analytical result of ref. \cite{ForRys},
which describes the asymptotic of a amplitude in the limit
$q^2>>(4m^2+Q^2)$, contains a uncertainty.  According to
eq.~(11) of ref. \cite{ForRys}, at $\rho>>1/Q$
\begin{equation}
\int d^2R J_0(QR){|\rho|\over |R+\rho/2||R-\rho/2|}\approx
J_0({1\over 2}QR){4\pi \over Q} \ .
\label{rys}
\end{equation}
This result is obtained if we anticipate that the main
contribution to the integral gives the small $\Delta R\sim 1/Q$
regions in the vicinity of the points ${\bf R}=\pm {\bf
\rho}/2$. Authors of ref. \cite{ForRys}
advocate that (\ref{rys}) confirms the "natural" conjecture that
in the limit $q^2>>(4m^2+Q^2)$ the contribution of the diagrams,
in which the $t-$ channel gluons connect with the various quarks
in a meson, can be neglected. And one can use the prescription
of ref. \cite{MuTa} for the amplitude of quark--quark scattering
to calculate the contribution of diagrams in which the both
gluons in the $t$-channel link to the same quark in a meson.

Note, that there are the other integration regions that give the
contribution to (\ref{rys}) of the same order as the discussed
above small region in the vicinity of the singular points.  The
contribution of the region $0\leq R \leq \rho^\prime
\stackrel{<}{\sim}
\rho/2$ can be estimated as $\approx {8\pi \rho^\prime \over
\rho Q}J_1(Q\rho^\prime)$.
The region $\rho^/2 \stackrel{<}{\sim}
\rho^{\prime\prime}\leq R \leq \infty$ gives the contribution
$\approx -{2\pi \rho \over \rho^{\prime\prime}
Q}J_1(Q\rho^{\prime\prime})$.  Because the function $J_1$
oscillates quickly when $Q\rho >>1$, one can not hold that
$J_1(Q\rho^\prime) \approx J_1(Q\rho^{\prime\prime})$.  Hence,
the contributions of these regions do not compensate each other
exactly and give the result of the same order as contribution
that gives the small regions in the vicinity of the points ${\bf
R}=\pm {\bf \rho}/2$.  Therefore, the asymptotic result of ref.
\cite{ForRys} for the amplitude
in the limit of high transfers $q^2>>(4m^2+Q^2)$ is only
parametricaly valid.  The numerical coefficient at the
asymptotic is uncorrect.  Nevertheless, the difference of true
numerical coefficient and that derived in ref. \cite{ForRys} is
presumable modest. Hence, authors of ref.
\cite{ForRys} do not detect this inaccuracy at the comparison
of the asymptotic expression with the results of numerical
calculations.

The integral of type (\ref{rys}) appears at the analysis of any
exclusive reaction in a frame of the BFKL theory.  As was
pointed out in ref. \cite{RLaK0}, the reduction of full integral
to the contribution of small regions near singular points is
equivalent to the approximation when we take into account only
diagrams, in which the $t$- channel gluons link to the same
parton, and use the prescription of ref. \cite{MuTa} for the
amplitude of quark--quark scattering.  We discussed in above
paragraph that in the region of high transfers this
approximation ((MT) approximation) gives right power behaviour
of the amplitude, but it is not sufficient to calculate exactly
numerical coefficient. Nevertheless, as the (MT) approximation
significantly simplifies the calculations, it can be used for
the estimate of an amplitude in a complicated situation, when
the exact calculation connects with the substantial
difficulties. This estimate can give the result which does not
differ tangibly from the exact one.  For instance, for the
discussed above process $\gamma q\to \rho^0 q$ (MT)
approximation gives the answer
\begin{equation}
A^{(MT)}(s,t)={32\over 3\pi^2}(1+{2\over 5}b_V)/(1+{131\over 320}b_V)
A(s,t) \ ,
\label{new}
\end{equation}
that exceeds the exact one (\ref{21}) less than on $10\%$
practically at any $b_V$.

\newpage

\medskip
\centerline{CAPTION}
\medskip

\noindent Fig.1.
One of the diagrams that describes the process of diffractive
vector meson production at large momentum transfer.  Other
diagrams are obtained by the all possible transpositions of
gluon's lines.

\medskip

\end{document}